\documentclass[12pt]{article}
\usepackage{epsfig,amssymb,amsfonts,newlfont,rotate}

\newcommand{\bbox}[1]{\mbox{\scriptsize\boldmath $#1$}} \bf
\begin{document}
\title{
Study of the Coulomb-Higgs transition\\ 
in the Abelian Higgs Model
}
\author{ A.~Cruz$^{a}$,
D.~I\~niguez$^{a}$,
L.~A.~Fern\'andez$^{b}$,\\
A.~Mu\~noz-Sudupe$^{b}$,
A.~Taranc\'on$^{a}$\\}
\bigskip
\maketitle

\begin{center}
{\it a)}  Departamento de F\'{\i}sica Te\'orica, Facultad de Ciencias,\\
Universidad de Zaragoza, 50009 Zaragoza, Spain \\
{\it b)} Departamento de F\'{\i}sica Te\'orica I, Facultad de Ciencias
F\'{\i}sicas,\\
Universidad Complutense de Madrid, 28040 Madrid, Spain \\
\end{center}
\bigskip
\begin{abstract}
The order of the Coulomb-Higgs transition in the U(1)-Higgs
model with unfrozen modulus of the scalar field is studied. 
Large lattices (up to $24^4$ in one case) and high statistics are used.
We fix $\beta =1.15$ and explore specially a region of $\lambda$-values where 
metastability is observed. 
We study the thermodynamical limit of several observables,
in particular, the latent heat, the specific heat, the decrement of the 
free energy between the maxima and the central minimum of the two-peaked 
histogram, the Binder cumulant and the displacement of the critical 
coupling with the lattice size. 
The results point towards a second order transition for 
$\lambda \gtrsim 0.005$,
while for smaller values of $\lambda$ the strong metastability growing with
the lattice size seems to derive from a first order character.
\end{abstract}

Monte Carlo.
U(1).
Higgs.
Phase transition.

\medskip


\newpage

\section{Introduction}

The Higgs mechanism is an essential part of the present day formulation of the 
Standard Model. The U(1) gauge model coupled to scalars is a simplification of 
the gauge--Higgs sector of the SM, which keeps the unsolved problem of
defining non perturbatively a non--asymptotically free field theory. 

The U(1)-Higgs model has been widely studied previously. One of the 
main objectives has been to determine the order of the Coulomb-Higgs 
transition, both in the cases of frozen and unfrozen modulus of the
scalar field, by using analytical (mean field, one-loop effective potential,
etc) and numerical techniques (see \cite{Espriu, Evertz87} and references
therein).

From those analyses the situation could be defined in the following way.
For large values of $\lambda$ the transition seems to be second order, while
for small values it looks first order. These results are based on Monte Carlo
studies of the system searching for metastabilities or single/double peaked 
histograms. Due to computational limitations, these calculations had been made 
with small lattices and short statistics. Here we carry out a study with
much larger lattices and high statistics in order to approach the 
thermodynamical limit in a more reliable way, obtaining results qualitatively
consistent with the previous ones.

However, in those works the conclusion that the transition is first order 
has been obtained
by considering the presence of a double peak for a given volume $V$ 
(or observing metastability). As we will show
this is not correct because even in this case, when $V$ is increased, both
peaks approach, and the latent heat disappears in the thermodynamical limit,
obtaining in this way a second order 
transition for $\lambda$ values much smaller than previously 
considered.

\section{The model and the Coulomb-Higgs transition}

The three parameter U(1)--Higgs model is described by the action
\begin{eqnarray}
S=-&\beta&\sum_{{\bbox r},\mu<\nu} \Re U_{{\bbox r},\mu\nu} \ 
-\ \kappa\sum_{{\bbox r},\mu} \Re\bar\Phi_{\bbox r}U_{{\bbox r},\mu}\Phi_{{\bbox r}+\mu}
\ +\cr
&\lambda&\sum_{\bbox r}(\vert\Phi_{\bbox r}\vert^2-1)^2\ +\ \sum_{\bbox r}\vert\Phi_
{\bbox r}\vert^2
\label{ACT3}
\end{eqnarray}
In the $\lambda\to\infty$ limit, $\vert\Phi\vert\to 1$ and the action 
simplifies to
\begin{equation}
S=-\beta\sum_{{\bbox r},\mu<\nu} \Re U_{{\bbox r},\mu\nu}
-\kappa\sum_{{\bbox r},\mu} \Re\bar\Phi_{\bbox r}U_{{\bbox r},\mu}\Phi_{{\bbox r}+\mu}
\label{ACT2}
\end{equation}

\begin{figure}
\begin{center}
\leavevmode
\centering\epsfig{file=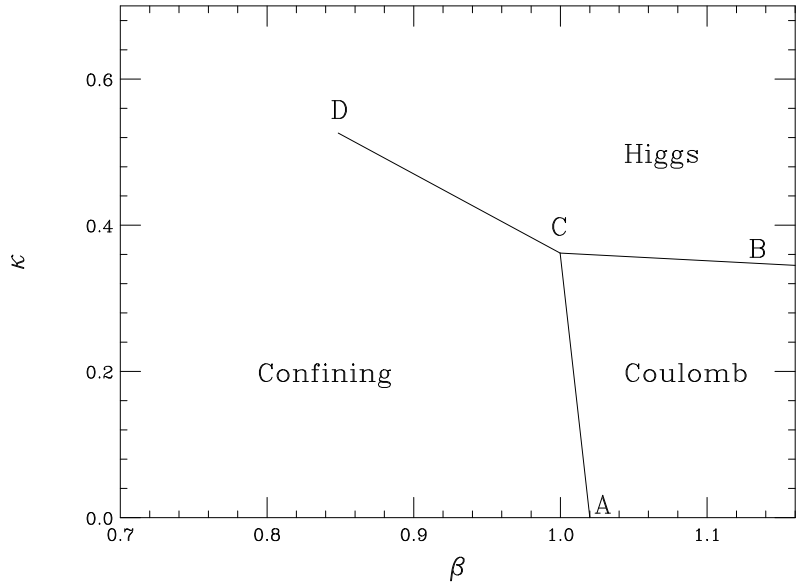,angle=90,width=300pt}
\end{center}
\caption{Phase diagram of the model with fixed modulus.}
\label{phdiu1}
\end{figure}

The phase diagram of that restricted version  was considered first in 
\cite{Fradk79} and has been
discussed by us in \cite{Cruz94}, and \cite{StLouis}. 
We discuss here the global aspects of the phase diagram in the fixed
modulus case (See Figure \ref{phdiu1}).
Point A is the pure compact $U(1)$ phase transition, a well established 
first order point \cite{Creutz},\cite{Lautrup},\cite{Petronzio}, even though
this transition for small lattices seems to be second order. We point out
that some authors \cite{Lang} have cast a new hue on its nature, pointing out
the possibility of this transition to be second order.

As we showed in \cite{Cruz94}, particularly in the neighbourhood of the triple 
point C, the line joining A with the latter is also first order. The line
CD was studied by us \cite{RTN_1} and is a first order line ending in a 
second order point D. The precise location of D is $\kappa=0.5260(9)$ and
$\beta=0.8485(8)$, with measured critical exponents compatible with the
classical (Mean Field) values $\alpha=0,\beta=1/2,\gamma=1,\nu=1/2$. 
The line BC is more controversial. 
The difficulty to identify
the order of the Coulomb--Higgs phase transition was made apparent because of
the large finite size effects.
The $\beta\to\infty$ end of that line is the X--Y model phase transition, 
a well established second order one with Mean Field exponents, 
which has induced researchers to think
the line to be a 
prolongation of its end point, and as such, second order \cite{Jansen86}. 
Yet, the relationship of the model to the $q>1$ version and to the
$Z_N$ model reported in \cite{Z8}, \cite{Cruz94} and \cite{StLouis} 
points towards a 
possible first order transition. However that relationship is based on
perturbative arguments (expansion around $\beta=\infty$) and might not be
applicable.

The difficulty to study directly the $\lambda\to\infty$ limit has lead us
to study the problem at finite, variable $\lambda$. This had been done earlier
\cite{Evertz87}, and we intend to improve on the statistics and the
lattice sizes.

\section{Simulation and observables}

We have fixed $\beta = 1.15$, which is on the Coulomb--Higgs side
of the first order Confining--Coulomb phase transition. 

If we consider larger values of $\beta$, the system has a larger correlation
length, and then in order to approach the thermodynamical limit, $L\gg\xi$, we
need very large volumes. Also, a tricritical point along the Coulomb--Higgs
transition, where the transition order changes, is not expected.

Then, we select some value of $\lambda$ and search for the ``critical'' 
$\kappa$, i.e. we are looking for the Coulomb-Higgs transition.
As is known, for small enough values of $\lambda$ the simulations on this
transition show clear signals of metastability, giving place to two-peaked
histograms, while for large values the metastability disappears. Of course, 
the simulations are made on finite lattices and then a conclusion on 
the order of the transition cannot be extracted directly from that
behaviour.
We have selected an intermediate region of $\lambda$ values, ranging from 
the point  where the two peak signal practically disappears for our 
lattices sizes ($\lambda=0.3$) to the point where the metastability is so 
strong that it makes the work with our computational disponibilities difficult 
($\lambda=0.003$). 
The total set of $\lambda$-values that we have used is 
$0.003, 0.005, 0.01, 0.03, 0.1, 0.3$ on lattices
$6^4, 8^4, 12^4, 16^4$ and $24^4$ (depending on $\lambda$), with statistics 
ranging between $10^5$ and $10^6$
Monte Carlo iterations per measurement, on workstations, on parallel machines
at the Edinburgh Parallel Computing Centre and on our custom, 64 T800
processor computer RTN \cite{CHEP92}.
We have used an over-relaxed Metropolis update method, with a multicanonical 
procedure to accelerate the flip-flop rate in some case.
We use the spectral density technique \cite{Ferr} in order to improve 
the determination of the transition point,
and the jack-knife method in the error estimation.

We have observed the different energies one can
define from the action. In particular we have paid special attention to
the link energy $E_\mathrm{link}=\langle\bar\Phi U\Phi \rangle$ as we have 
fixed the value 
of $\beta$ and the transition is almost parallel to the $\beta$ axis.   
All the observables we will define hereafter will be referred to this 
energy.

\begin{figure}
\begin{center}
\leavevmode
\centering\epsfig{file=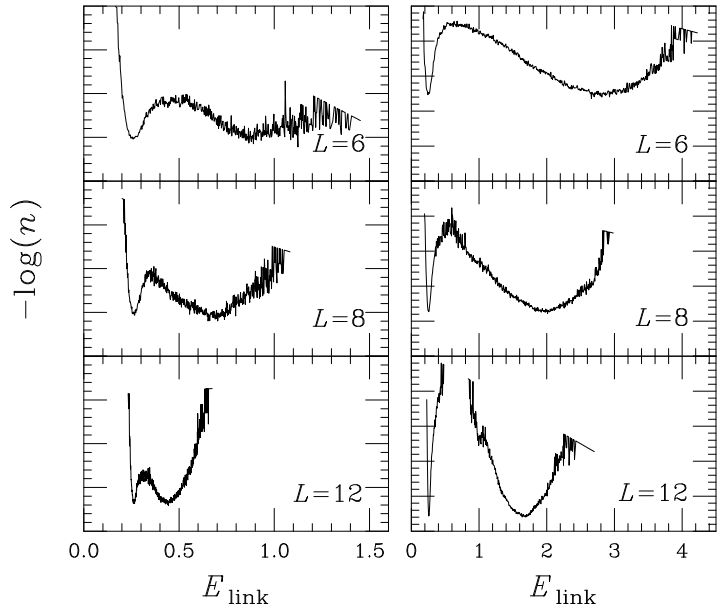,width=230pt,angle=90}
\end{center}
\caption{Minus logarithm of the $E_\mathrm{link}$ histogram at
$\lambda=0.01$ (left) and  $\lambda=0.003$ (right).}
\label{free_energy}
\end{figure}

In the range of $\lambda$ and $L$ studied, the histograms of $E_\mathrm{link}$
present the following features (examples can be seen 
in fig. \ref{free_energy}):

\begin{itemize}
\item Two-peak structure.
\item Asymmetry of the peaks, with a narrow low-energy peak and a broad 
high-energy one.
\item Strong size dependence, consisting on a narrowing of the gap and width
of the peaks as the lattice size $L$ increases, caused by a much faster 
displacement towards lower energy of the broad peak.
\end{itemize}

It is known that, due to finite size effects, two-peaked histograms on
finite lattices can become single peaked in the thermodynamical limit
and vice versa. For instance, the transition in the $q=4$ Potts model
is known analytically to be second order but the simulation on finite
lattices gives a double peaked structure \cite{Fukugita}.
On the contrary, single peaked
histograms on small lattices can become double peaked when the lattice
size increases, as in \cite{ising4d}.

Then an accurate study of finite size effects has to be made. We will use 
several independent Finite Size Scaling methods, which allow to determine 
correctly
the order of the transition whenever one uses large enough lattices (but
not so large as if one only looked for metastability signals). Now we will
introduce 
the different quantities which we have considered in order to extract our
conclusions.

First, coming back to the histograms, in ref. \cite{Lee} an interesting 
method is proposed. They explain how in a two-peaked structure the 
increment between the minima and the local maximum of the free energy
is directly related to the analogous increment for the 
logarithm of the histogram of the energy. 
 The idea is that if the transition  
is first order the gap has to be more pronounced for larger $L$, while if
it is second order the depth of the gap does not grow with $L$. Our procedure
will be the following. From the histogram of $E_\mathrm{link}$ obtained in the 
simulation, we
calculate a new histogram where the height of the two peaks is the same
by using basically the spectral density method \cite{Ferr}. 
Then we take the logarithm
of it and change its sign. In this way we obtain a figure analogous to the
free energy (see fig. \ref{free_energy}). Here we can measure directly the
depth of the gap, let us call it $\Delta F$. A $\Delta F$ growing with $L$
indicates a first order transition, while if it is constant we have
a second order one. 

Second, we have measured the specific heat
\begin{equation}
c_V = \frac{\partial \langle E_\mathrm{link} \rangle}{\partial \kappa}.
\end{equation}
Its maximal value should scale with an exponent $\alpha/\nu$. It 
will be $d$ for a first order transition and $0$ for a mean field
second order one (in this case $c_V^{\mathrm max}$ will be a constant plus 
logarithmic corrections).

Third, we use the maximum of $c_V$ as definition of {\em critical}
kappa $\kappa_\mathrm{c}(L,\lambda)$.  The scaling of
$\kappa_\mathrm{c}(L)$ with $L$ allows us to extract $\nu$. It will be
$\nu=1/d$ if the transition is first order, and $\nu=1/2$ if we have
second order with mean field exponents.

Fourth, we consider the latent heat
at $\kappa_\mathrm{c}(L,\lambda)$,  $\Delta E$, as 
the  difference 
between the positions of the maxima for a fit to each peak separately 
to a cubic spline, after the histogram has 
been shifted to the apparent critical point by the spectral density method.
 A cubic spline has been preferred to other functional shapes
because of its ability to reproduce a maximum and accommodate the mixed states, 
whose influence on the histogram is otherwise difficult to account for. If
we have $\Delta E\ne 0$ in the $V\to\infty$ limit we have a first order
phase transition.

Fifth, we have computed the Binder cumulant
\begin{equation}
B=1-\frac{\langle E_\mathrm{link}^4 \rangle}{3{\langle
E_\mathrm{link}^2 \rangle}^2}.
\end{equation}
If the energy distribution is single peaked (and then second order) in the 
thermodynamical limit, this
quantity can be computed exactly and becomes $2/3$.

\section{Results}

First of all, let us comment on which have been our limits on
$\lambda$ and $L$. 

For the biggest value of $\lambda$ studied, $\lambda=0.3$, it was impossible 
to distinguish the two peaks on the histogram, and then we could not make
the type of treatment described above. For $\lambda=0.1$ two peaks are visible
but the gap is so small that it is difficult to measure with precision
observables such as $\Delta F$ and $\Delta E$; for this value of $\lambda$
we have only simulated $L=6,8$. 

On the other hand, for the smallest value
$\lambda=0.003$ the metastability is so strong that we have only been
able to use $L=6,8,12$. Besides, the simulation in the $L=12$
have been made fixing a value $\kappa=0.2703$ and making two different 
runs, one starting from a cold configuration and the other one starting from
a hot one. In this case ($L=12$) no flip was observed; then we can 
only say that the two peaks are rather separated, but a precise estimation
of the latent heat and of the other observables is not possible. In
particular, the latent heat is greatly affected by a displacement in 
$\kappa$ (because the broad peak is much more sensitive than the narrow one,
as the displacement in the peak will be proportional to the squared width) 
and a bad estimation 
of $\kappa_\mathrm{c}$ would give place to an erroneous measurement.

For the intermediate values of $\lambda$ things are more feasible and we have
used $L=6,8,12,16$ for $\lambda=0.005,\ 0.01,\ 0.03$ and also $L=24$ for
$\lambda=0.005$.

\begin{figure}
\begin{center}
\leavevmode
\centering\epsfig{file=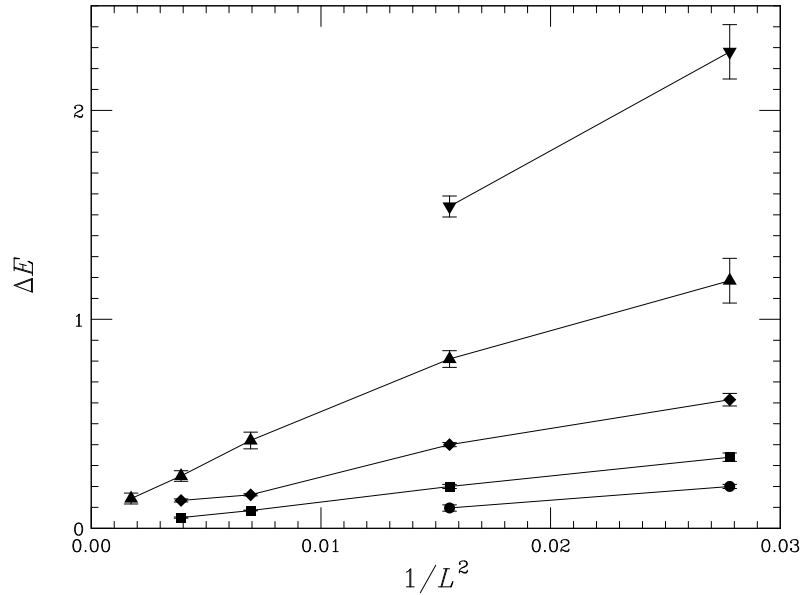,angle=90,width=300pt}
\end{center}
\caption{Latent heat as a function of $1/L^2$ at different values of $\lambda$.
From top to bottom, $\lambda=0.003,0.005,0.01,0.03,0.1$.}
\label{latent}
\end{figure}

Let us now present the results for the different observables.

The study of the latent heat suggests an extrapolation to zero in the
thermodynamical limit for $\lambda \gtrsim 0.005$ as shown in
fig. \ref{latent}.

The specific heat $c_V$ is constant with $L$ for the largest $\lambda$-values
while it grows for the smallest ones, as shown in table \ref{table_heat}. 
In the presence of a stable double peaked structure (i.e. such that the peaks
do not approach when changing $L$), $c_V$ scales with the maximum possible
speed, i.e. like $V$, indicating a first order transition. However in our
case the peaks are not stable and then the scaling will be different. For
a trivial second order transition in $d=4$, it scales only logarithmically
($\alpha=0$), then it must be almost constant. The results indicate a 
crossover between this two behaviors at $\lambda\approx 0.005$.   

\begin{table}[t]
\begin{center}
\footnotesize
\begin{tabular}{|c||c|c|c|c|c|}
\hline
$\lambda$ & $L=6$ & $L=8$ & $L=12$ & $L=16$ & $L=24$ \\
\hline
\hline
0.003 & 7710(490) & 9810(1150) &  &  & \\
\hline
0.005 & 1820(80) & 2400(70) & 3850(170) & 4260(890) & 7760(810)\\
\hline
0.01  & 577(57) & 652(53) & 660(28) & 897(105)&\\
\hline
0.03  & 185(41) & 189(17) & 172(15) & 186(12)&\\
\hline
0.1   & 63(2) & 58(3)&&&\\
\hline
\end{tabular}
\normalsize
\caption{Specific heat for the different values of $L$ and $\lambda$.} 
\label{table_heat}
\end{center}
\end{table}

\begin{figure}
\begin{center}
\leavevmode
\centering\epsfig{file=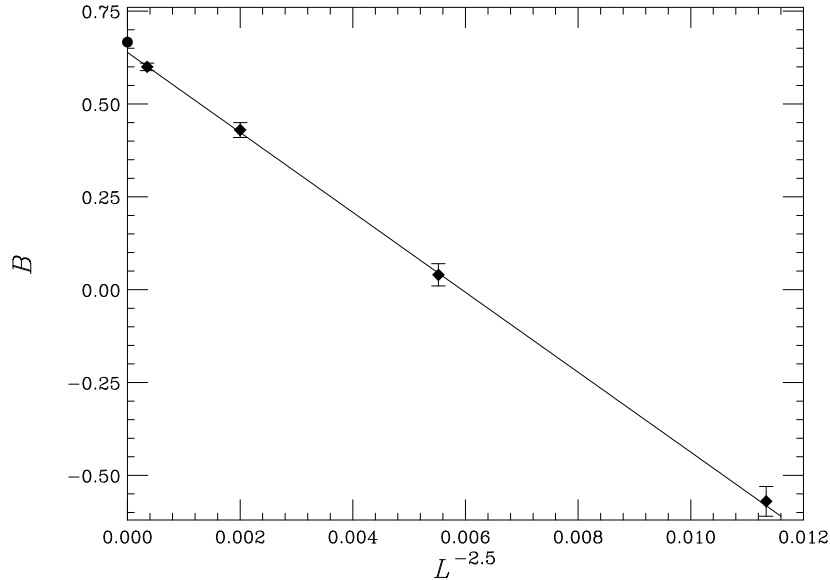,width=220pt,angle=90}
\end{center}
\caption{Binder cumulant at $\lambda=0.005$ as a function of $L^{-2.5}$.
The circle corresponds to the value $2/3$.}
\label{binder}
\end{figure}

For the values of $\lambda$ where we have more measurements, i.e. 
$\lambda=0.005$, $0.01$, $0.03$ we have fitted $\kappa_\mathrm{c}(L)$ as 
a function of $L$:
\begin{equation}
\kappa_\mathrm{c}(L)=\kappa_\mathrm{c}(\infty)+AL^{-1/\nu}.
\end{equation}

The three parameter fit gives $\nu=0.40(1),0.43(1),0.44(1)$ respectively
at those values of $\lambda$. These values of $\nu$ are closer
to the mean field one, $\nu=1/2$, than to the expected in a first
order transition, $\nu=1/4$. However, those apparent 
$\nu$-exponents would probably
derive to one of the two 
values (1/2 or 1/4) in the thermodynamical limit.
   
The behaviour of the Binder cumulant has not given any conclusive result.
In principle, it should approach the value 2/3 with corrections of the order
of $L^{2(1/\nu-d)}$ for a second order transition or a value different from
2/3 with corrections as $L^{-d}$ for a first order one. In practice, we 
obtain some value close to 2/3 but neither the precision
is enough nor the scaling is as cited. In fig. 
\ref{binder} we show the results for $\lambda=0.005$. We have a good 
scaling but with an exponent $-2.5(2)$ instead of $-4$, which
would be expected both in the first order case and in the gaussian second
order one. The fit gives a value $0.64(1)$ in the infinite volume limit.
However this difference from 2/3 is not significant as the error is large
and the scaling is still not as expected in the thermodynamical limit.

\begin{table}[!t]
\begin{center}
\footnotesize
\begin{tabular}{|c||c|c|c|c|c|}
\hline
$\lambda$ & $L=6$ & $L=8$ & $L=12$ & $L=16$ & $L=24$ \\
\hline
\hline
0.003 & 4.0(4) & 5.0(7) & large &  & \\
\hline
0.005 & 2.3(5) & 2.1(4) & 3.1(4) & 3.4(4) & 3.5(5)\\
\hline
0.01  & 1.7(4) & 1.9(4) & 1.3(4) & 1.9(4)&\\
\hline
0.03  & 0.6(2) & 0.8(3) & 0.6(2) & 0.7(2)&\\
\hline
\end{tabular}
\normalsize
\caption{$\Delta F$ for the different values of $L$ and $\lambda$.} 
\label{table_libre}
\end{center}
\end{table}

Finally, in table \ref{table_libre} are shown the increments of the free energy
$\Delta F$. Again we can see that the behavior changes at
$\lambda\approx 0.005$; for larger values $\Delta F$ is constant with $L$,
while for smaller ones it grows. A look at fig. \ref{free_energy} 
shows the different
kind of behavior for $\lambda=0.01$ and $\lambda=0.003$; while in the first
case the gap tends to disappear when increasing the lattice size, 
in the second one it tends to become deeper.

\section{Conclusions}

We have investigated the order of the Coulomb-Higgs transition at fixed 
$\beta=1.15$ and for different values of $\lambda$.
The analysis
of several observables indicates a crossover at $\lambda\approx 0.005$,
the transition being first order for smaller and second 
order for larger values of $\lambda$.
We remark that in order to extract this conclusion we have needed to consider
different lattice sizes and, even though most of them present a two-peak 
structure, the latent heat goes to zero in the thermodynamical limit 
(at least for $\lambda \gtrsim 0.005$).
Previous works had estimated this crossover at $\lambda\approx0.3$ 
\cite{Jansen86} or $\lambda\approx0.02$ \cite{Espriu} for $\beta=1.5$.
However for larger $\beta$ a weaker character is expected (the correlation
length grows and we approach the second order transition of the 
$\phi^4$ theory).
Then for $\beta=1.5$ the crossover should occur at a smaller $\lambda$-value
than for $\beta=1.15$. Therefore we have lowered the estimation of the
values of $\lambda$ for which the Coulomb-Higgs transition changes order.

\section*{Acknowledgments}

We are very grateful to the RTN collaboration for the use of their 
transputer machine where we have run most of our simulations.
D.I. acknowledges the EPCC and the Deparment of Physics of Edinburgh,
as well as the Ministerio de Educaci\'on y Cultura (Spain) for a 
fellowship. This work has been partially supported by CICYT under contract 
numbers AEN96-1634 and AEN96-1670 and by the EU under contract number
ERB-CHRX CT 92-0051.

\end{document}